\documentclass[preprint,prb,aps,showkeys]{revtex4}
\usepackage{epsfig}
\usepackage{amsmath}
\begin{document}
\title{Anomalies in water as obtained from computer simulations of the TIP4P/2005 model:
density maxima, and density, isothermal compressibility and heat capacity 
minima}
\author{Helena L. Pi, Juan L. Aragones, Carlos Vega\footnote{cvega@quim.ucm.es}, 
Eva G. Noya, Jose L. F. Abascal, Miguel A. Gonzalez and Carl McBride} 
\affiliation{Departamento de Qu\'{\i}mica F\'{\i}sica,
Facultad de Ciencias Qu\'{\i}micas,\\ Universidad Complutense de Madrid,
28040 Madrid, Spain\\}

\date{\today}

\begin{abstract}
The so-called thermodynamic anomalies of water form an integral part of the peculiar
behaviour of this both important and ubiquitous molecule.
In this paper our aim is to establish whether the recently proposed TIP4P/2005 model
is capable of reproducing a number of these anomalies.
Using molecular dynamics simulations we investigate both the maximum in density 
and the minimum in the isothermal compressibility along a number of isobars.
It is shown that the model correctly describes the decrease in the temperature
of the density maximum with increasing pressure.
At atmospheric pressure the model exhibits an additional minimum in density at a
temperature of about 200K, in good agreement 
with recent experimental work on super-cooled confined water.
The model also presents a minimum in the isothermal compressibility
close to 310K.  
We have also investigated the atmospheric pressure isobar for three other water models;
the SPC/E and TIP4P
models also present a minimum in the isothermal compressibility, 
although at a considerably lower temperature than the experimental one.
For the temperature range considered no such minimum is found for the TIP5P model.
\end{abstract}
\maketitle

\vspace{0.5cm}

\section{Introduction}
The study of polar fluids has long been a topic of 
interest,  even more-so since the development of perturbation theories and computer 
simulation techniques\cite{weis1,weis2}. Amongst the many polar molecules, 
water stands out in particular. 
Water is a fascinating molecule, 
both from a practical and from a fundamental
point of view.  
In the liquid phase water presents a number of anomalies when compared to other liquids
\cite{eisenberg69,bookPhysIce,finneyreview,chaplin,JML_2001_90_0303},
whilst the solid phase exhibits a highly complex phase diagram,
having at least fifteen different crystalline structures \cite{eisenberg69,iceXIII}. 
Due to its importance and its complexity, understanding the properties 
of water from a molecular point of view is of considerable interest and presents a veritable challenge. 

Computer simulations of water started with the pioneering papers by Watts and 
Barker \cite{CPL_1969_3_0144_nolotengo} and by Rahman and 
Stillinger \cite{JCP_1971_55_03336} about forty years ago. 
However, a key issue that still exists when performing simulations of water is the choice of model for the pair potential
that is used to describe the interaction
between molecules \cite{lynden,laaksonen,ferrarioKF,paschek,waterinproteins}. 
The SPC\cite{berendsen82}, SPC/E\cite{berendsen87}, TIP3P\cite{jorgensen83}, TIP4P\cite{jorgensen83} 
and TIP5P \cite{mahoney00} models have become highly popular among the large community of 
people simulating water or water solutions. 
Each of the aforementioned models are rigid and non-polarisable, 
but naturally, real water is both flexible and can be polarised. It is almost needless to say
that these models
represent a very simplified version of the true interactions between real water 
molecules. 
Water also exhibits important quantum nuclear effects, so a 
realistic description of water should also take this into account \cite{guillot02}. 
That said, it is of interest at this stage to analyse how far it is possible to
go in  describing  real water using these simple models. 

In recent years the inexorable increase in computing power has permitted 
the calculation of properties that were previously inaccessible to simulations.
These properties can be used as new `target quantities'
when fitting the  parameters for any new potential.
More importantly, some of these properties have provided  stringent
tests for the existing water models. 
In particular, some of the authors have recently determined  the phase diagram for different
water models and have found that the performance of the models can vary significantly
\cite{sanz1,vegautrecht,vega_review,abascal05a,abascal05b}.
For example, it has been  found that TIP4P provides a qualitatively correct description of the 
global phase diagram of water, whereas the SPC, SPC/E, and TIP5P models do not.\cite{vega_faraday}
For the SPC and SPC/E models  the melting temperature and the maximum in density 
of water occur at temperatures far below the experimental values.
Taking this into account some of the authors proposed a new rigid non-polarisable model
with the following constraints:
that it should be based on the TIP4P model, since this model provides a reasonable 
description of the phase diagram of water. 
It should also reproduce the maximum in density of liquid water (notice that most of
models fail when it comes to predicting the  location of the density maximum \cite{vega05b}).
Finally, the model should account for the vaporisation enthalpy of real water, 
but only after incorporating the self-polarisation energy correction  
proposed by Berendsen \emph{et al.} for the SPC/E model \cite{berendsen87}. 
It was from these considerations that the TIP4P/2005 model of water arose.\cite{abascal05b}
In the paper in which the model was presented it was shown that TIP4P/2005
correctly describes  the global phase diagram of water, the diffusion coefficient at 
atmospheric pressure and temperature, the maximum in density along the atmospheric pressure 
isobar, the density of the ice polymorphs, and the equation of state of liquid
water at high pressures. 
Afterwards, additional studies have shown that the model is also able to 
provide a good description of the vapour-liquid equilibria\cite{vega06} and the surface 
tension\cite{vega07a}. 
Further information concerning the performance of TIP4P/2005 and its comparison
with other water models can be found elsewhere \cite{vega_faraday}.
The overall results indicate that TIP4P/2005 is probably close to being the best rigid non-polarisable model that  
can be achieved; any significant improvement 
would  require the introduction of `new' physics, i.e., flexibility, 
polarisability and nuclear quantum effects. 

In this paper we are interested in the thermodynamic response functions in the region where 
water exhibits `anomalous' behaviour.
In particular, these are the expansion coefficient (which vanishes at 
temperatures close to the melting point, resulting in the well known maximum
of density at about 4$^\circ$C at atmospheric pressure) and the isothermal 
compressibility \cite{debenedetti03} ($\kappa_T$ ) which shows a minimum at 46.5$^\circ$C at
$p$=1~bar.
Moreover, we shall investigate whether the model is able to describe the
pressure dependence of these thermodynamic properties, namely,
the decrease in the temperature of the density maximum and the shift towards 
slightly higher temperatures of the minimum in $\kappa_T$ as the pressure increases.
It will be shown that the model is indeed able to describe these two features quite 
well, which provides further evidence of the robust behaviour of the model even 
when estimating properties that were not taken into account in the original fitting process. 

\section{Methodology}
The interaction between water molecules will be described by the TIP4P/2005
model.\cite{abascal05b}  
In this model, a Lennard-Jones centre is located on the oxygen atom,
positive charges are placed on the hydrogen atoms and a negative charge is located at
the site $M$ situated along the H-O-H bisector.
For the simulations we have used the molecular dynamics package GROMACS 
(version 3.3).\cite{gromacs33}
The Lennard-Jones potential has been truncated at 9.0~\AA. 
Long range corrections were applied to the Lennard-Jones part of the potential (for both 
the energy and pressure).
Ewald summations were used to deal with electrostatic contributions.
The real part of the Coulombic potential was truncated at 9.0~\AA.
The Fourier component of the Ewald sums was evaluated by using the Particle Mesh
Ewald (PME) method of Essmann \emph{et al.}\cite{essmann95}
The width of the mesh was 1~\AA\ and a fourth degree polynomial was used. 
The simulation box was cubic throughout the whole simulation  and the geometry of the water
molecules was enforced using constraints.\cite{shake,berendsen84}
The temperature was set by using a Nos{\'e}-Hoover\cite{nose84,hoover85} 
thermostat with a relaxation time of 2~ps.
To maintain constant  pressure an isotropic Parrinello-Rahman
barostat\cite{parrinello81,nose83} with a relaxation time of 2~ps was used. 
As a check, at  two pressures, $p$=1~bar and $p$=400~bar, Monte
Carlo simulations were performed using a bespoke program. 
The Monte Carlo  densities were in complete agreement with those obtained from molecular dynamics using GROMACS.

To determine the maximum in density, molecular dynamics simulations have been performed along 
the isobars $p$=1, 400, 1000, and 1500 bar. 
The number of molecules used in the simulations was 256. 
Long runs are required to determine the maximum in density; 
for each thermodynamic state twenty million time-steps were performed. 
Since the time step was 2~fs, the results presented here are an average of 
the properties of the system obtained from runs of 40~ns. 
The simulations were started at high temperatures, and the final configuration 
of a particular run was used as the initial configuration for a lower temperature
simulation.
Typically about 6 to 8 different temperatures were studied along each isobar.
The isothermal compressibility is defined as:
\begin{equation}
\label{eqkappa}
  \kappa_T=-\frac{1}{V}\left(\frac{\partial V}{\partial p}\right)_T.
\end{equation}
The literature for the isothermal compressibility of water models is
rather scant\cite{motakabbir90,glattli02}.
The computational overhead required for 
a study of the variation of $\kappa_T$ with temperature with sufficient accuracy
has, until now, been prohibitive.
Here we have  evaluated the isothermal compressibility for the following water models: 
SPC/E\cite{berendsen87}, TIP4P\cite{jorgensen83}, TIP4P/2005\cite{abascal05b}
and TIP5P\cite{mahoney00} along the atmospheric pressure isobar.
The isothermal compressibility has been evaluated using the well known 
fluctuation formula
\begin{equation}
\label{eqfluct}
  \kappa_T=\frac{\left<V^2\right>-\left<V\right>^2}{k T \left<V\right>}.
\end{equation}
The volume fluctuations were typically averaged over 20~ns using a time-step
of 1~fs.
Preliminary results for TIP4P/2005 model were sufficiently close to the experimental
value as to warrant a more precise calculation. 
Thus, for this model, the isothermal compressibility has been determined via two 
different procedures using, in both cases, a time-step of 0.5~fs and a sample size
of 500 molecules.
The first method is the application of the volume fluctuation formula, typically over 
40 million molecular dynamics time-steps, for a total simulation length of about 20ns.
The relative uncertainty of the calculated compressibilities is about 
$\pm 3$\%.
The second route is the numerical evaluation of the derivative appearing in 
the definition of $\kappa_T$ (Eq.~\ref{eqkappa}). 
To this end, the results for the equation of state of five different
state points (at different pressures but at the same temperature) were fitted
to a second degree polynomial.
In addition to the state point of interest, for which the volume is already known from the runs used to calculate
$\kappa_T$ via the fluctuation route, 
four additional runs are required:
two runs at the same temperature but at pressures of 200 and 400 bar higher,
and another two runs at pressures 200 and 400 bar lower. 
The simulation length of these additional runs was 6.5 ns , so in the end the
computational cost of both routes is almost the same.
We have also computed the heat capacity at constant pressure $C_p$ for 
the TIP4P/2005 model. $C_p$ is defined as : 
\begin{equation}
\label{eqcp}
  C_p=\left(\frac{\partial H}{\partial T}\right)_p.
\end{equation}
Thus in order to compute $C_p$ the enthalpy at each temperature was first calculated.  
These values for the enthalpy were then fitted to a polynomial and
this fit was then differentiated with respect to temperature to obtain $C_p(T)$.

\section{Results}
\begin{figure}
\includegraphics[clip,width=0.6\textwidth,angle=-0]{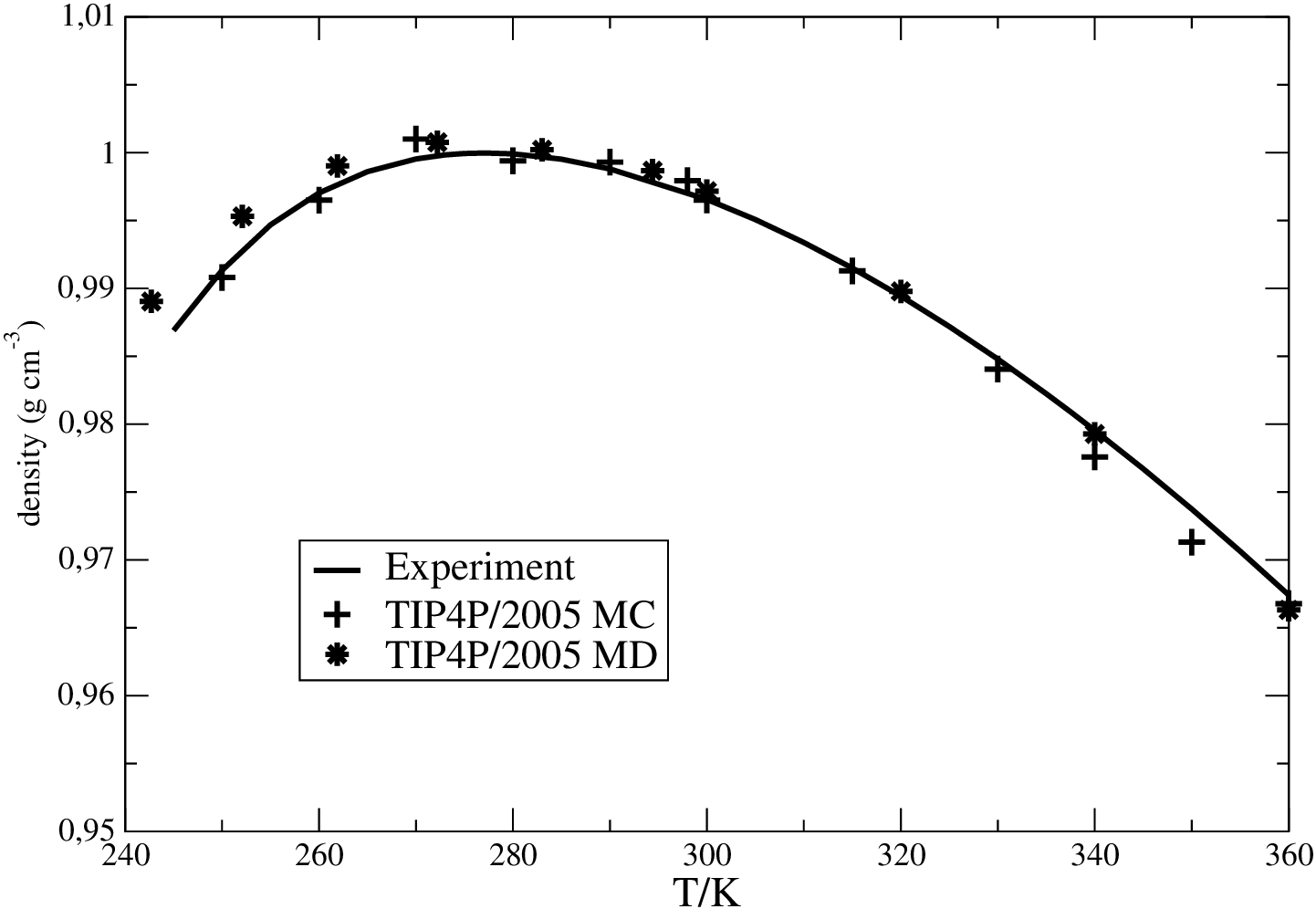}
\includegraphics[clip,width=0.6\textwidth,angle=-0]{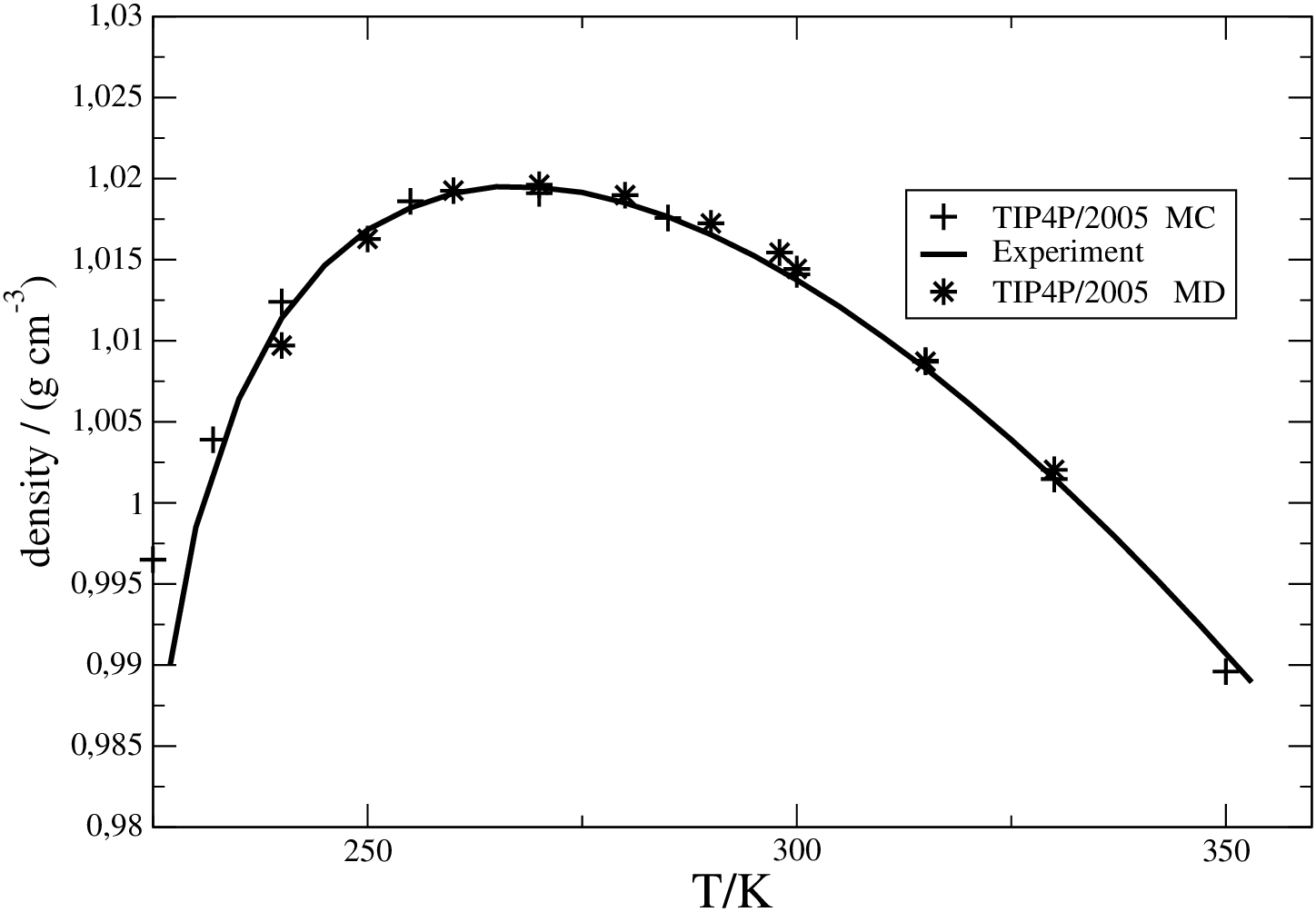}
\caption{\label{400barmcmd} Density of liquid water at $p=1$~bar (upper panel) 
and $p$=400~bar (lower panel) as obtained from Monte Carlo  and molecular dynamics simulations with the 
TIP4P/2005 model. 
The experimental results were taken from Ref. \onlinecite{JPCRD_1989_18_1537}.}
\end{figure}

\begin{figure}[!hbt]\centering
\includegraphics[clip,width=0.6\textwidth,angle=-0]{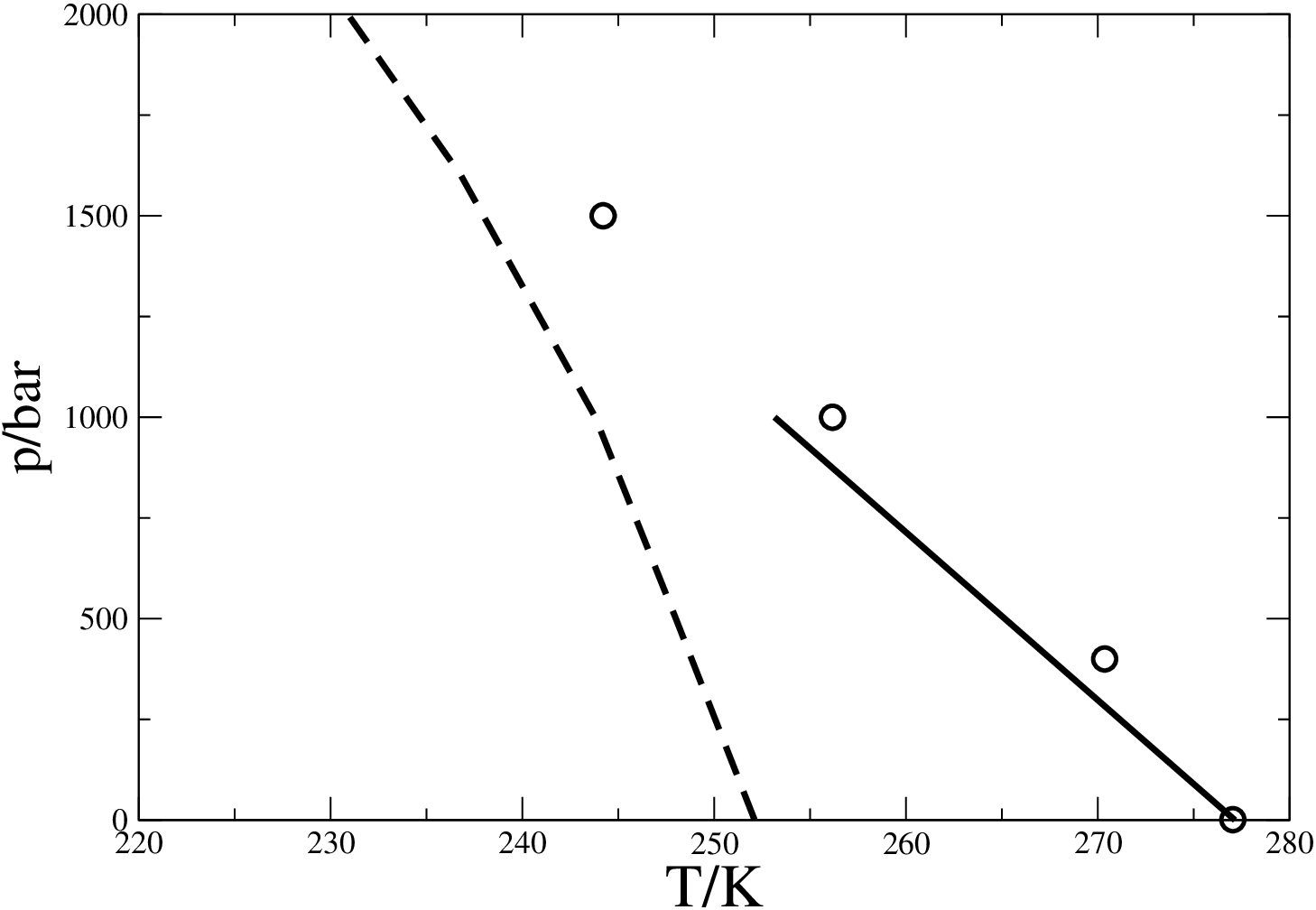}
\caption{\label{TMDfinal} Dependence of the TMD with pressure for the 
TIP4P/2005 model (open circles). Experimental data for the TMD (solid line) 
taken from Fig.8 of Ref.~\onlinecite{fine}. The dashed line is the melting 
curve of ice I$_h$ for the TIP4P/2005 model}.
\end{figure}

The simulation results for the densities are reported in Tables 
\ref{tip4p_2005_1bar}, \ref{tip4p_2005_400bar}, \ref{tip4p_2005_1000bar} 
and \ref{tip4p_2005_1500bar} for  $p$=1, 400, 1000 and 1500 isobars, 
respectively.
For the four  isobars considered a clear maximum in density has been found. 
The results for $p$=1~bar and $p=$400~bar, along with experimental measurements,
are plotted in Fig.~\ref{400barmcmd}. 
It can be seen that the model is able to reproduce the experimental
data quite nicely. 
The density maximum for each isobar was obtained by fitting the densities 
around the maximum (typically 5 or 6 densities were used in the fit) to a second 
or a third degree polynomial. 
The values of the maximum are reported in Table \ref{tablaTMD}. 
In Fig.~\ref{TMDfinal} the temperatures at which the maxima appear (TMD) are 
compared to the experimental values\cite{fine}. 
As can be seen, the agreement is rather good. 
The model correctly predicts a decrease of about 33~K in the temperature of the 
maximum when going from atmospheric pressure to a pressure of about 1500~bar.
In other words, 45 bar are required to induce a decrease in the TMD of about 
one degree. 
For D$_2$O Angell and Kanno found a similar lowering of the TMD with 
pressure.\cite{tmd_d2O} 
For the TIP4P/2005 not only does the TMD decrease with pressure 
but the melting temperature does too. 
In Fig.\ref{TMDfinal} the melting curve of ice I$_h$ (taken from our previous work\cite{abascal05b,pccp_coexistence_lines,pccp_plastic}) is plotted along the TMD curve determined in this work. 
As can be seen, both curves have a negative slope. Notice also that at room pressure
the distance between the melting curve of ice I$_h$ and the TMD is about 25K, which 
is in contrast with the 4K of difference found experimentally. The impossibility of 
reproducing simultaneously both the TMD and the melting temperature in models having three 
charges has been discussed previously\cite{vega05b,vega_faraday}. 

\begin{figure}[!hbt]\centering
\includegraphics[clip,width=0.6\textwidth,angle=-0]{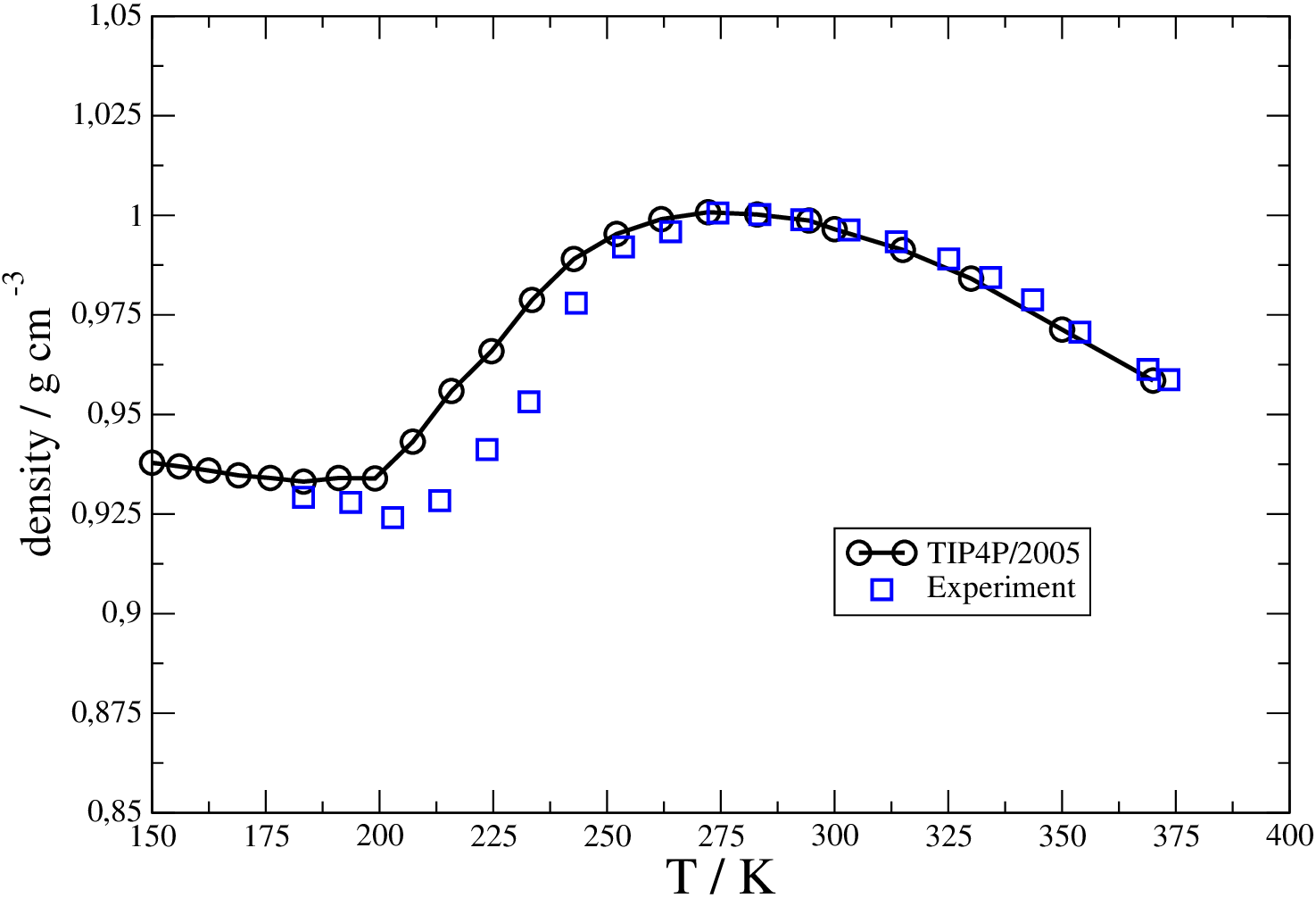}
\caption{\label{lahostia} Density of liquid water at atmospheric pressure
as obtained from molecular dynamics simulations with the TIP4P/2005 model.
For comparison, experimental data of water confined in narrow  pores
are  also given \cite{mallamace01}.}
\end{figure}

\begin{figure}[!hbt]\centering
\includegraphics[clip,width=0.6\textwidth,angle=-0]{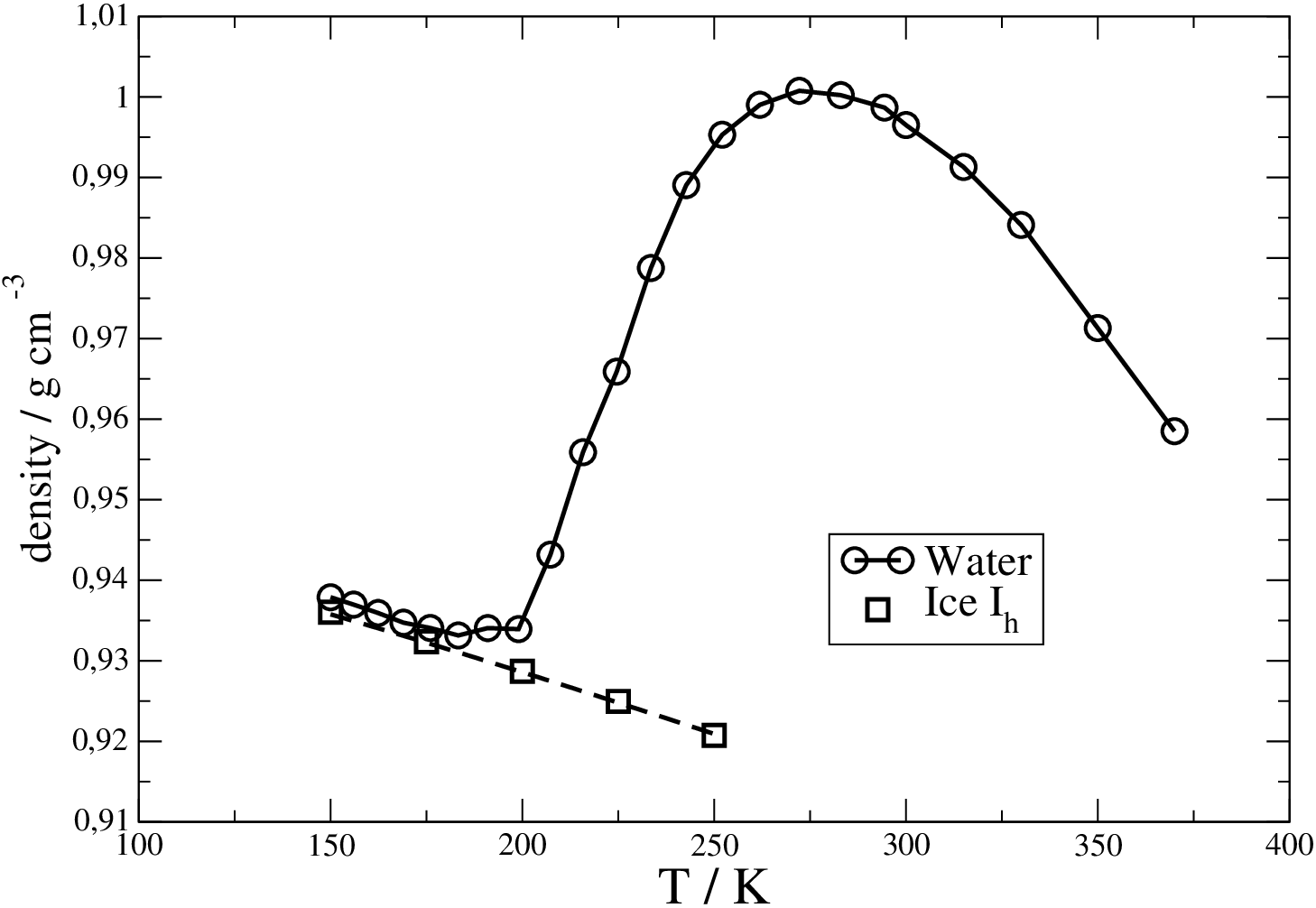}
\caption{\label{lahostia_2} Density of liquid water at atmospheric pressure
as obtained from molecular dynamics simulations with the TIP4P/2005 model (solid 
line and open circles). Density of ice I$_h$ at atmospheric pressure
as obtained from molecular dynamics simulations with the TIP4P/2005 model (dashed 
line and open squares). }
\end{figure}

Another interesting issue is the behaviour of the density along the atmospheric 
pressure isotherm at very low temperatures. 
Experimentally, it is not possible to obtain the density of water at 
temperatures below 233~K (the homogeneous stability limit of water at atmospheric 
pressure\cite{th}) due to spontaneous nucleation and freezing. 
However, recently, it has been possible to avoid the formation of ice by 
confining water in pores a few nanometers in diameter (most probably because of the
decrease in the freezing point of water due to confinement as described 
qualitatively by the Gibbs-Thomson approach\cite{gibbs_thomson}).
Thanks to this the density of deuterated water has  recently been determined for
the first time\cite{pnas_confined} for temperatures as low as 160~K, reporting
the existence of a minimum in the density at a temperature of about 200~K.
A similar study (for water instead of deuterated water) was performed 
by Mallamace \emph{et al.}\cite{mallamace01}. 
By using infra-red (IR) spectroscopy, it was possible to determine the density of liquid
water in the super-cooled regime down to 150~K.
In Fig.~\ref{lahostia} the reported experimental densities (from IR spectroscopy
of a sample within a 1.5~nm pore) for liquid water along the atmospheric pressure isobar
are compared to those obtained in this work from molecular dynamics simulations of the 
TIP4P/2005 water model.
The agreement is surprisingly good, and the location of the minimum around 
200~K is described properly by TIP4P/2005. Such a minimum in density has also been 
found\cite{paschek_minimum,sciortino_minimum} in computer simulations of the TIP5P and ST2 models of water,
but the agreement with experiment was not as good as that exhibited by TIP4P/2005. 
For temperatures below this minimum the density increases as the temperature 
decreases (as in a normal fluid). In Fig.~\ref{lahostia_2} we compare the equation of state 
of supercooled water to the equation of state of ice I$_{h}$ for the TIP4P/2005 model  
along the room pressure isobar.\cite{aragones2}
The minimum in density of supercooled water occurs 
just when the density  is approaching that of the ice I$_{h}$. However, we did not succeed in obtaining 
ice I$_{h}$ by cooling water; the radial distribution functions 
of the supercooled water (at room pressure) below 200K are clearly different from those of ice I$_{h}$. 
Rather the minimum in density corresponds to the formation of  a glassy state.     
Note that the existence of such a minimum is not only restricted to  water but
is also present in materials such as tellurium \cite{highs_and_lows}.
Interestingly the melting curve for tellurium exhibits re-entrance .
Such re-entrant behaviour was also found in our studies of the phase diagram of 
water models \cite{sanz1}.

\begin{figure}[!hbt]\centering
\includegraphics[clip,width=0.6\textwidth,angle=-0]{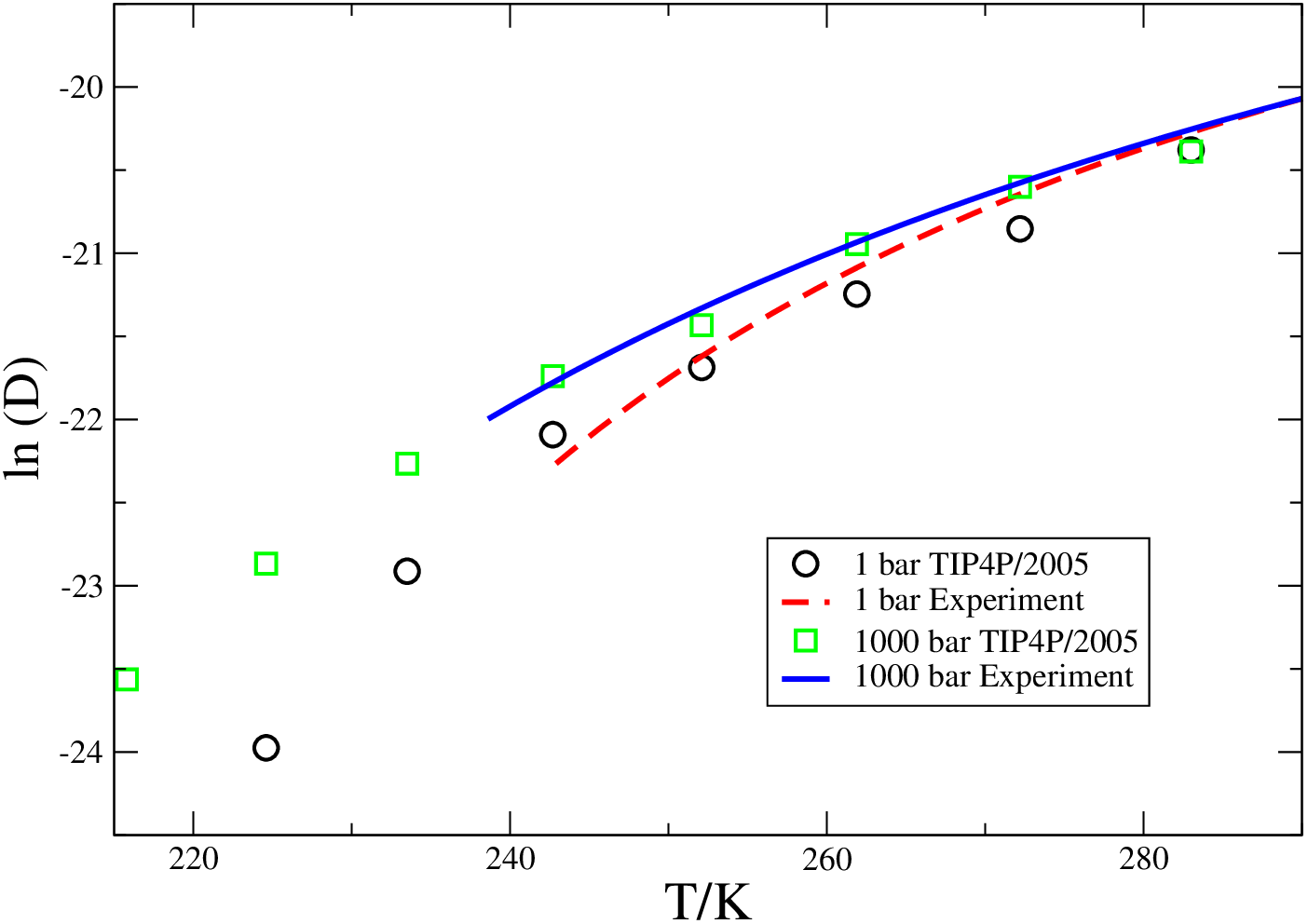}
\caption{\label{difusion} Self-diffusion coefficient (in cm$^2$/s) of the 
TIP4P/2005 model at $p$=1~bar and $p$=1000~bar compared to the experimental
results (taken from Ref.~\onlinecite{cdifusion}).}
\end{figure}

We have also computed the self-diffusion coefficient at $p=$1~bar and 
$p$=1000~bar.
The results, presented in Fig.~\ref{difusion}, show that the diffusion 
coefficient drops significantly as the temperature decreases. 
The decay is less pronounced at higher pressures. 
At low temperatures the diffusion coefficient increases significantly with 
pressure (this behaviour is anomalous\cite{debenedetti03} since, for most 
fluids, it decreases with increasing pressure). 
One can imagine that the application of pressure somehow breaks the hydrogen bond network, 
thus aiding diffusion processes. 
  Some indirect  evidence of the decrease in hydrogen bonding with 
  pressure can be obtained from the analysis of the different 
  contributions  (Lennard-Jones and Coulombic) to the residual internal 
  energy.  In Table \ref{contributions} 
  the different contributions to the residual internal 
  energy obtained from molecular dynamics  of the TIP4P/2005 model 
  along the T=224.6K isotherm  (at different pressures) are given. 
  Notice the large decrease (in absolute value) of the Coulombic energy
  with pressure, clearly pointing  to a reduction of 
  hydrogen bonding with pressure. The Lennard-Jones contribution is more repulsive at low pressures 
(even though the density is lower) as a consequence of the stronger hydrogen bond found 
 at lower pressures. 
At temperatures above 280~K the diffusion coefficients at the two pressures considered 
 become virtually identical as the differences fall to within the 
statistical uncertainty. The agreement with the experimental values of the 
diffusion coefficient\cite{cdifusion} is quite good. 

\begin{figure}[!hbt]\centering
\includegraphics[clip,width=0.49\textwidth,angle=0]{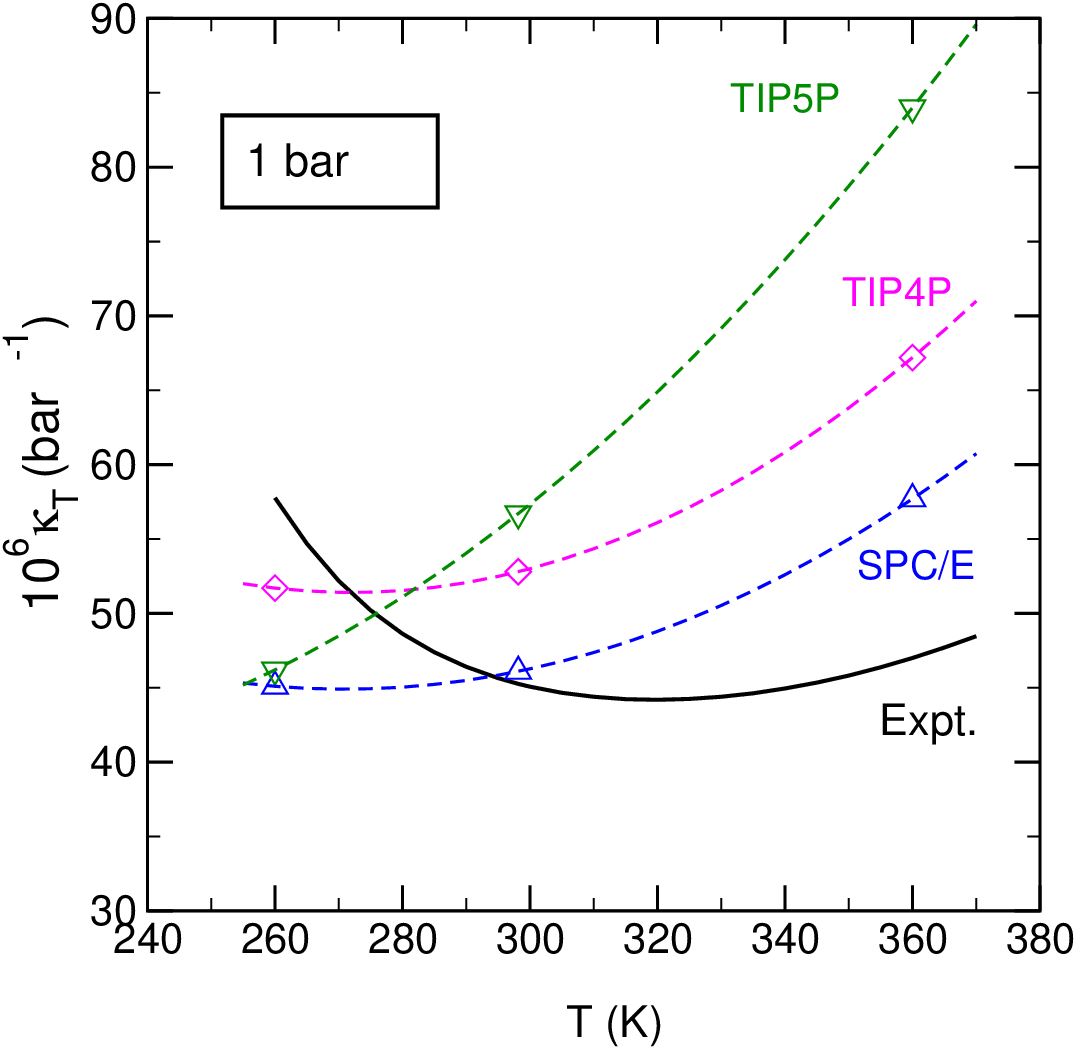}
\includegraphics[clip,width=0.49\textwidth,angle=0]{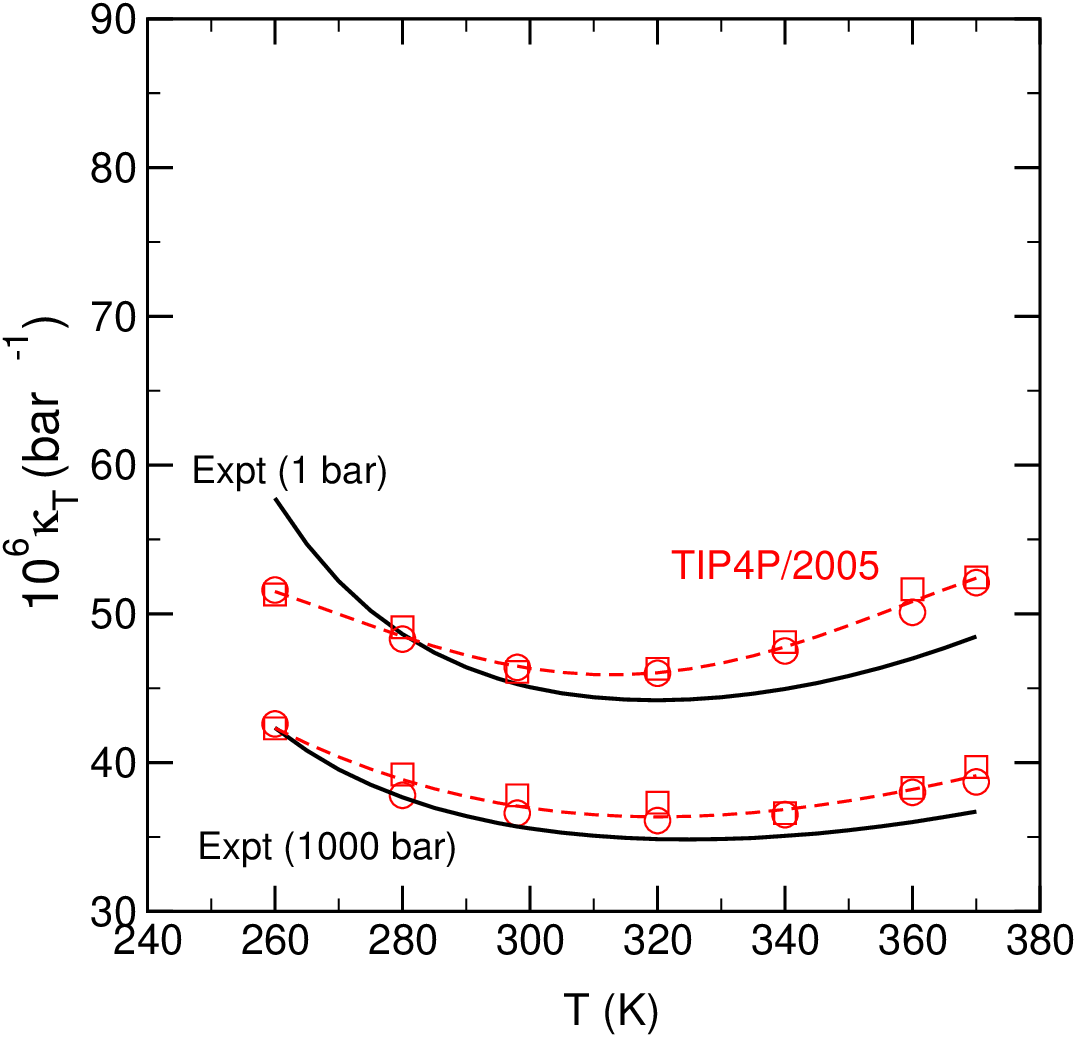}
\caption{\label{compresibilidad} Isothermal compressibility calculated 
from molecular dynamics simulations.
Left: Results for SPC/E, TIP4P and TIP5P at $p$=1~bar using the fluctuation 
formula (Eq.~\ref{eqfluct}).
Right: Results for TIP4P/2005 using the fluctuation equation (squares) and 
the derivative route (Eq.~\ref{eqkappa}, circles); 
upper results are for 1~bar and lower curves are for 1000~bar.
Notice that we have deliberately used the same scale in both panels.
For comparison, experimental data\cite{JPCRD_1989_18_1537} (full lines) are 
also plotted.}
\end{figure}

Let us now focus on another of the `anomalous' properties of water, the 
isothermal compressibility.
The experimental measurements\cite{kell75,speedy76,JPCRD_1989_18_1537,carlton07} 
show that, at atmospheric pressure, $\kappa_T$ drops 
as the temperature increases from the melting temperature up to 46.5$^\circ$.
Above this temperature water behaves as a normal liquid and the isothermal
compressibility increases with temperature.
It is also well established from experiment that the temperature for which
$\kappa_T$ is minimal shifts slightly towards higher values as the pressure increases.
Contrary to the maximum in density, the ability of the water
models to account for the compressibility minimum has not yet been established.
As mentioned in the previous section,  simulation studies of the isothermal
compressibility are few and far between\cite{motakabbir90,glattli02}.
Moreover, the computational resources available did not allow the extended
simulations needed to calculate $\kappa_T$ with the precision required to
determine  whether the most common water models predict the compressibility
minimum. 
For this reason we have calculated the compressibility for a select few  `popular'
water models.
The results are presented in Table \ref{kt_modelos} for the SPC/E, TIP4P and TIP5P models 
and in Table \ref{kt_bio7} for TIP4P/2005. 
The calculations for SPC/E, TIP4P and TIP5P were performed  at atmospheric pressure
for three different temperatures, namely 260 K, 300 K and 360 K.
Fig.~\ref{compresibilidad}a shows that the temperature dependence of the
isothermal compressibility for TIP5P does not follow the experimental pattern 
since, for this model,  $\kappa_T$ is a monotonously increasing function along
the whole experimental liquid range.
Despite the fact that the slope of the TIP5P curve is opposite to the experimental
one, the values of $\kappa_T$ are coincident in a narrow range of temperatures
because the simulation and the experimental curves cross at a temperature 
close to the freezing point of liquid water.
For temperatures near the boiling point the TIP5P model
fails completely, where  the predicted compressibility at 360 K is almost twice the
experimental value.

As for the performance of the SPC/E and TIP4P models, both provide fairly similar
results.
In fact their curves are parallel, showing a more or less defined minimum around
270 K (more computations would be needed to determine the precise location of 
the minima).
The results of SPC/E are somewhat shifted to higher temperatures with respect
to those of TIP4P, which results in a slightly better agreement with the
experimental data.
At high temperatures, the differences between simulation and experiment are 
quite noticeable for both SPC/E and TIP4P (though less dramatic than in the 
TIP5P case).

As can be seen in Fig.~\ref{compresibilidad}b, the TIP4P/2005 model provides
an excellent description of the isothermal compressibility of water.
The compressibilities obtained from the two routes (differenciation of the equation
of state and fluctuation formula) were found to be mutually consistent. 
The departures of the calculated compressibilities with respect to the 
experimental values\cite{kell75,speedy76,JPCRD_1989_18_1537,carlton07} are in general smaller than 7\%.
By fitting the TIP4P/2005 compressibilities at atmospheric pressure to a third
degree polynomial, a minimum in the isothermal compressibility is found for a 
temperature of about 37$^\circ$C which is in  good agreement with the 
experimental value (46.5$^\circ$C).
Thus the model is able to accurately reproduce not only the isothermal 
compressibility along the whole liquid range at atmospheric pressure but also 
the compressibility minimum.
This assertion is particularly true when one compares the TIP4P/2005 predictions
with those of the other models, shown in Fig~\ref{compresibilidad} 
(notice that we have plotted both panels using the same scale).
In Fig~\ref{compresibilidad}, we also present the compressibility results at
a higher pressure (1000~bar).
TIP4P/2005 predictions for this isobar are  slightly better than those for
atmospheric pressure.
A compressibility minimum is also found at 1000 bar, and, in accordance with 
the experiment, the minimum appears at an slightly higher temperature than it does at
atmospheric pressure \cite{JPCRD_1989_18_1537}.

\begin{figure}[!hbt]\centering
\includegraphics[clip,width=0.6\textwidth,angle=-0]{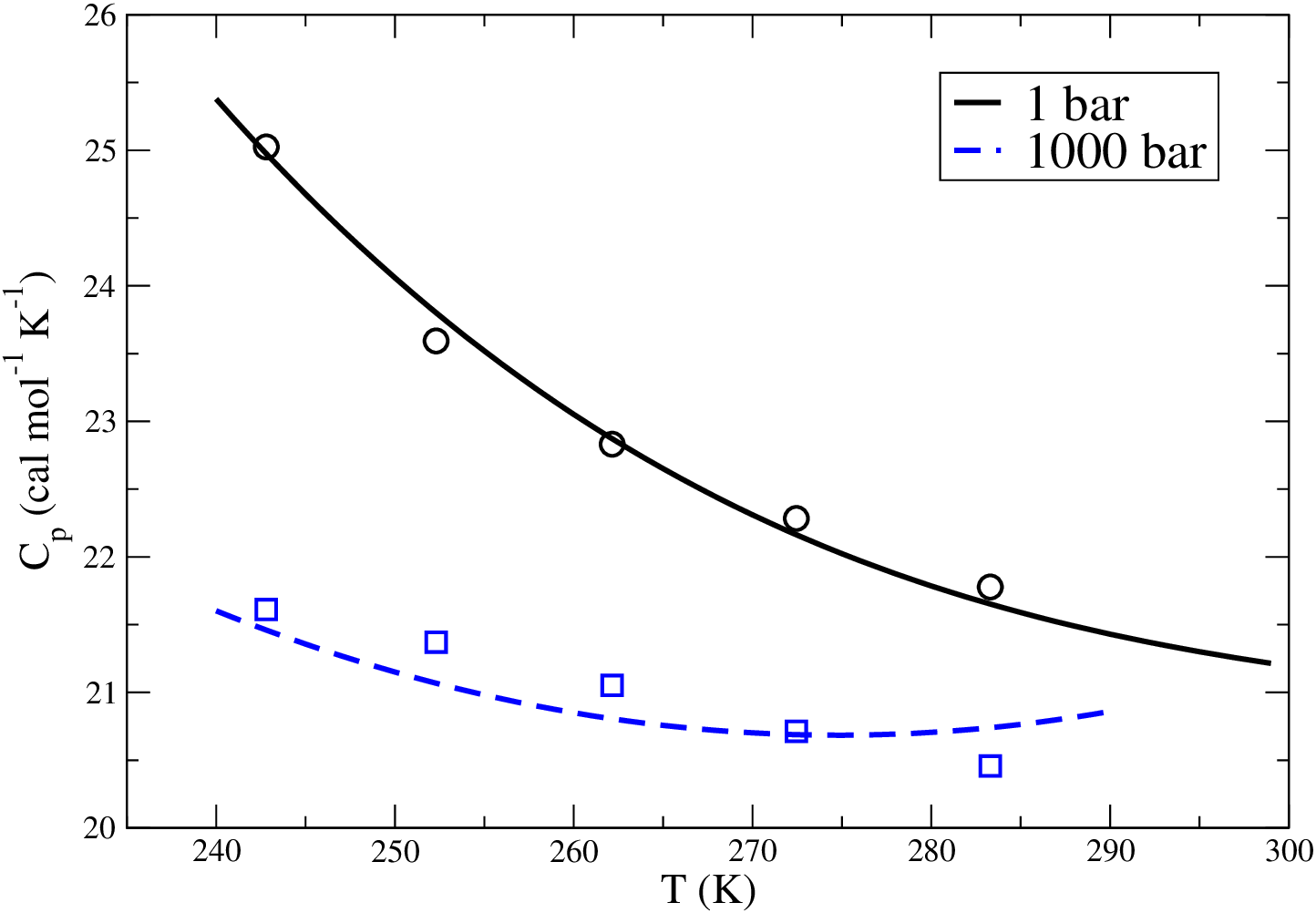}
\caption{\label{heat_capacity} Heat capacity of water at constant pressure 
($C_p$) as obtained from simulation results of the TIP4P/2005 for 
$p=$ 1 bar (solid line/circles) and for $p=$ 1000 bar (dashed line/squares).
The symbols represent the results of a first degree finite difference calculation for 
equation (3), and the curves were obtained from the results of a third degree polynomial fit (p=1000 bar ) and fourth degree (p=1 bar) for the enthalpy. }
\end{figure}

 Finally we have examined the behaviour of the heat capacity at constant
pressure for two isobars, namely, $p=$ 1 bar and $p=$ 1000 bar. The results
are presented in figure \ref{heat_capacity}. As can be seen, the model 
hints at  the existence of minima in the heat capacity for 
both of these isobars.
The location of the minimum seems to move to lower temperatures as the pressure 
increases, in concordance with experiment (see fig.7 of Ref.\cite{fine}).

\begin{figure}[!hbt]\centering
\includegraphics[clip,width=0.6\textwidth,angle=-0]{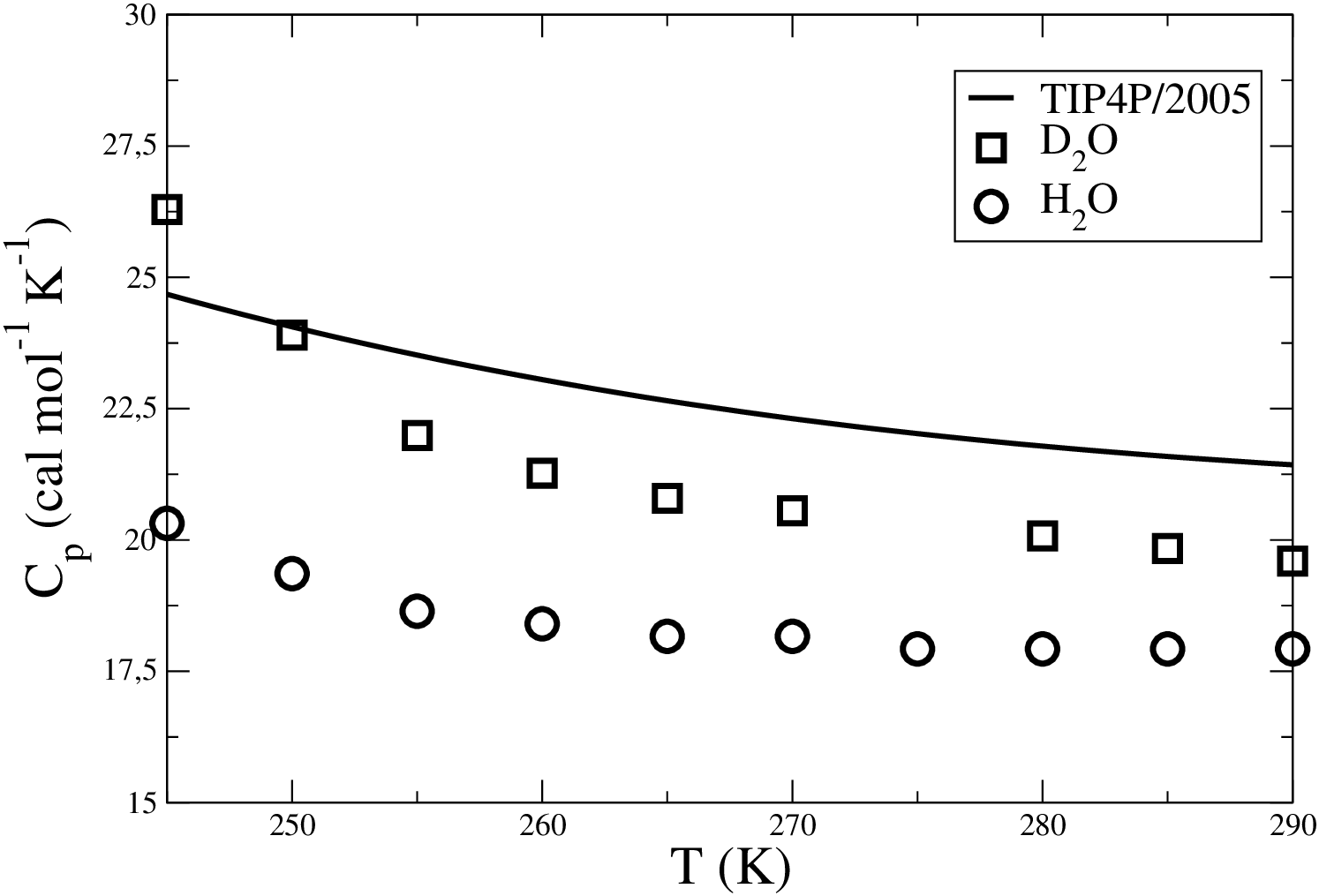}
\caption{\label{cp_md_vs_exp} Heat capacity of water at constant pressure 
( $C_p$ ) as obtained from simulation results of the TIP4P/2005 for 
$p=$ 1 bar . Experimental results for D$_2$O and for H$_2$O taken from Angell,Oguni and 
Sichina \cite{cp_h2o_d2O_exp} are also presented.}
\end{figure}

In figure \ref{cp_md_vs_exp} the  values of $C_p$  from the simulations  are compared 
to the experimental\cite{cp_h2o_d2O_exp} 
values for water and D$_2$O. As one can see, the 
model overestimates the experimental values of the heat 
capacity of water at constant pressure. The large difference in the 
experimental values of the heat capacity of H$_2$O and D$_2$O demonstrate
the importance of nuclear quantum effects in the description of the 
heat capacity of water. Not surprisingly, the values of TIP4P/2005 
which were obtained through classical simulations lie closer to 
the experimental values of D$_2$O, probably reflecting the somewhat 
more classical behaviour of D$_2$O with respect to H$_2$O.

Taking into account the success of TIP4P/2005 in  describing 
most of the properties
of water, the failure in the quantitative description of $C_p$ (along 
with the important differences in the experimental values of D$_2$O and 
H$_2$O) points out the necessity of incorporating nuclear quantum 
effects for a quantitative description of this property. Although one could incorporate
quantum corrections empirically\cite{horn04,abascal05b}, quantum simulations (for example, via path integrals) are probably the only way to 
 correctly describe the heat capacity of water\cite{quantum_1,quantum_2}.

\section{ Conclusions } 
In this work molecular dynamics simulations have been undertaken for liquid water along several
isobars.
The TIP4P/2005 model has been used to describe the interaction between water 
molecules.  
The results obtained in relatively long runs (i.e., 40~ns) indicate that
a maximum in density is found for the isobars considered in this work 
(i.e., $p$=1, 400, 1000 and 1500~bar). 
The TMD decreases by about 1 K for each 45~bar of applied pressure. 
Thus, the TIP4P/2005 model, designed to reproduce the TMD at atmospheric pressure, 
is also able to  predict the dependence of the TMD with pressure. 
Motivated by recent experimental work the behaviour of the
density at low temperatures along the atmospheric pressure isobar was also studied, 
resulting in a density minimum at temperatures around 200~K.
The location and density at the minimum of the model are in very nice agreement
with recent experimental work on the equation state of water at very low 
temperatures (obtained by confining water in narrow  pores in order to prevent the
nucleation of ice). 
Also in agreement with experimental measurements we have found a significant 
increase of the diffusion coefficient of super-cooled water with pressure. 

Finally, we have also computed the isothermal compressibility along the  atmospheric
pressure isobar for several water models and, once again,  the results for TIP4P/2005 
are those closest to experimental values.
A minimum in the isothermal compressibility was found for temperatures around 
37$^\circ$C, once again in  good agreement with the location of the minimum found in 
experiments (46.5$^\circ$C) .
Not only is the location of the minimum well described by the model, but the 
value at the minimum as well. 
Finally, we have calculated $\kappa_T$ at several temperatures at a higher 
pressure (1000~bar) observing that TIP4P/2005 also exhibits a compressibility
minimum.
In accordance with experiment, the minimum is shifted towards slightly higher 
temperatures\cite{fine1973} when the pressure increases 
from 1 bar to 1000 bar. 

The results of this work indicate that the TIP4P/2005 is able to reproduce 
almost quantitatively many of the anomalous properties of water occurring 
at low temperatures with the exception of the heat capacity. Therefore, the model 
can be used with confidence to obtain a better understanding of the behaviour of
water in the super-cooled regime where the experimental determination of properties 
is a difficult task. 
It seems that for super-cooled water the performance of the TIP4P/2005 model
of water supersedes the 
performance of other more traditional models. 
 
\acknowledgments
This work has been funded by grants FIS2007-66079-C02-01
from the DGI (Spain), S-0505/ESP/0299 from the CAM, 
and 910570 from the UCM.

\bibliographystyle{./apsrev}
\bibliography{./bibliography_review}

\begin{center}
\begin{table}[!h]
\begin{center}
\caption{Molecular dynamics results for the TIP4P/2005 model of water 
along the $p=$ 1 bar isobar.
Only the residual part of the internal energy is given.}
\label{tip4p_2005_1bar}
\begin{tabular}{cccc}
\hline \hline
 T/K & p/bar & $\rho$/(g/cm$^3$) & U/(kJ/mol) \\
\hline

150.0 & 1 & 0.9379 & -57.05 \\
156.0 & 1 & 0.9370 & -56.87 \\
162.4 & 1 & 0.9359 & -56.68 \\
169.0 & 1 & 0.9347 & -56.47 \\
176.0 & 1 & 0.9341 & -56.25 \\
183.3 & 1 & 0.9331 & -55.99 \\
191.0 & 1 & 0.9341 & -55.74 \\
199.0 & 1 & 0.9339 & -55.51 \\
207.3 & 1 & 0.9432 & -54.81 \\
215.8 & 1 & 0.9559 & -54.04 \\
224.6 & 1 & 0.9659 & -53.27 \\
233.5 & 1 & 0.9787 & -52.47 \\
242.7 & 1 & 0.9890 & -51.70 \\
252.1 & 1 & 0.9953 & -50.98 \\
261.9 & 1 & 0.9991 & -50.28 \\
272.2 & 1 & 1.0008 & -49.57 \\
283.0 & 1 & 1.0002 & -48.84 \\
294.4 & 1 & 0.9987 & -48.10 \\
300.0 & 1 & 0.9972 & -47.73 \\
\hline \hline

\end{tabular}
\end{center}
\end{table}
\end{center}

\newpage 

\begin{table}[!ht]
\begin{center}
\caption{\label{tip4p_2005_400bar} 
Molecular dynamics results for the TIP4P/2005 model of water along the $p$=400 bar isobar.
Only the residual part of the internal energy is given.}
\begin{tabular}{cccc}
\hline \hline
 T/K & p/bar & $\rho$/(g/cm$^3$) & U/(kJ/mol)  \\
\hline 
215.8 & 400 &  0.9766 & -53.39  \\
224.6 & 400 &  0.9963 & -52.92  \\
233.5 & 400 &  1.0051 & -52.27  \\
242.7 & 400 &  1.0125 & -51.57  \\
252.1 & 400 &  1.0171 & -50.90  \\
261.9 & 400 &  1.0192 & -50.24 \\
272.2 & 400 &  1.0200 & -49.55  \\
283.0 & 400 &  1.0190 & -48.85  \\
294.4 & 400 &  1.0164 & -48.14  \\
\hline \hline

\end{tabular}
\end{center}
\end{table}

\newpage 

\begin{table}[!hb]
\begin{center}
\caption{\label{tip4p_2005_1000bar}
Molecular dynamics results for the TIP4P/2005 model of water along the $p$=1000 bar isobar.
Only the residual part of the internal energy is given.}
\begin{tabular}{cccc}
\hline \hline
 T/K & p/bar & $\rho$/(g/cm$^3$) & U/(kJ/mol) \\
\hline
215.8 & 1000 & 1.0308 & -53.25 \\
224.6 & 1000 & 1.0397 & -52.62 \\
233.5 & 1000 & 1.0438 & -52.04 \\
242.7 & 1000 & 1.0463 & -51.43 \\
252.1 & 1000 & 1.0476 & -50.82 \\
261.9 & 1000 & 1.0475 & -50.19 \\
272.2 & 1000 & 1.0462 & -49.55 \\
283.0 & 1000 & 1.0441 & -48.89 \\
294.4 & 1000 & 1.0407 & -48.20 \\
\hline \hline

\end{tabular}
\end{center}
\end{table}

\newpage 

\begin{table}[!hp]
\begin{center}
\caption{\label{tip4p_2005_1500bar}
Molecular dynamics results for the TIP4P/2005 model of water along the $p$=1500 bar isobar.
Only the residual part of the internal energy is given.}
\begin{tabular}{cccc}
\hline \hline
 T/K & p/bar & $\rho$/(g/cm$^3$) & U/(kJ/mol) \\
\hline
199.0 & 1500 & 1.0472 & -54.25 \\
207.3 & 1500 & 1.0617 & -53.61 \\
215.8 & 1500 & 1.0667 & -53.06 \\
224.6 & 1500 & 1.0678 & -52.52 \\
233.5 & 1500 & 1.0702 & -51.94 \\
242.7 & 1500 & 1.0701 & -51.37 \\
252.1 & 1500 & 1.0700 & -50.78 \\
261.9 & 1500 & 1.0682 & -50.18 \\
272.2 & 1500 & 1.0662 & -49.56 \\
283.0 & 1500 & 1.0631 & -48.91 \\
294.4 & 1500 & 1.0593 & -48.24 \\
\hline \hline

\end{tabular}
\end{center}
\end{table}

\newpage 

\begin{table}[!h]
\begin{center}
\caption{\label{tablaTMD}
Temperature of maximum density at different pressures. 
The TMD has been obtained by fitting the data in the proximity
of the maximum to a quadratic or a cubic polynomial. 
The temperature ranges used in the fit were  260-300~K for 
$p$=1~bar, 242-295~K for $p$=400~bar, 233-283~K for $p$=1000~bar, and 207-272~K
for $p$=1500~bar.}
\begin{tabular}{cccccccccc}
 \hline \hline
p/bar & & & 1      & & 400    & & 1000   & & 1500 \\
TMD/K & & & 277(3) & & 270(3) & & 256(3) & & 244(3) \\
\hline \hline
\end{tabular}
\end{center}
\end{table}

\newpage 
\begin{table}[!h]
\begin{center}
\caption{\label{contributions}
  Different contributions to the residual internal energy 
  of water along the $T=224.6K$ isotherm. Results were obtained 
  from molecular dynamics simulations of 
  the TIP4P/2005 model of water. The Lennard-Jones contribution  
  $U_{\mathrm{Lennard-Jones}}$ , the Coulombic contribution $U_{\mathrm{Coulombic}}$ and the total 
  residual energy $U$ are reported. 
  All energies are given in kJ/mol. }  
\begin{tabular}{cccc}
 \hline \hline
     p/bar      &   $U_{\mathrm{Lennard-Jones}}$ &  $U_{\mathrm{Coulombic}}$ &     $U$           \\
    1    &    11.34  &     -64.61  &     -53.27     \\
    400  &    11.08  &     -64.00  &     -52.92    \\
    1000 &    10.78  &     -63.40  &     -52.62     \\
    1500 &    10.66  &     -63.18  &     -52.52    \\
\hline \hline
\end{tabular}
\end{center}
\end{table}

\newpage 

\begin{table}[!h]
\begin{center}
\caption{\label{kt_modelos}
Isothermal compressibility $\kappa_{T}$ for $p=$ 1 bar as obtained in this work for 
the TIP4P, SPC/E and TIP5P models of water. 
Reported values correspond to ($\kappa_{T}/$ bar$^{-1}$) $\times$10$^{6}$.}
\begin{tabular}{ccccc}
 \hline \hline
T/K           &  TIP4P    &   SPC/E   &  TIP5P & Experiment \\         
260 &    51.8     &  45.1      &         46.5  &    57.8    \\
298.15 & 52.8     &  46.1      &         57    &    45.3    \\
360 &    67.2     &  57.7      &         84    &    47.0     \\
\hline \hline
\end{tabular}
\end{center}
\end{table}

\newpage 

\begin{table}[!h]
\begin{center}
\caption{\label{kt_bio7}
Isothermal compressibility $\kappa_{T}$ for $p=$ 1 bar and $p=$ 1000 bar 
as obtained in this work for the TIP4P/2005 model of water. 
Reported values correspond to ($\kappa_{T}/$ bar$^{-1}$) $\times$10$^{6}$.}
\begin{tabular}{cccc}
 \hline \hline
T/K & p/bar &    TIP4P/2005  &    Experiment \\
260 & 1     & 51.4 & 57.8 \\
280 & 1     & 48.7 & 48.6 \\
298 & 1     & 46.3 & 45.3 \\
320 & 1     & 46.2 & 44.2 \\
340 & 1     & 47.8 & 44.9 \\
360 & 1     & 50.9 & 47.0 \\
370 & 1     & 52.3 & 48.5 \\
260 & 1000  & 42.4 & 42.3 \\
280 & 1000  & 38.5 & 37.7  \\
298 & 1000  & 37.2 & 35.7 \\
320 & 1000  & 36.7 & 34.9 \\
340 & 1000  & 36.6 & 35.1 \\
360 & 1000  & 38.2 & 36.0 \\
370 & 1000  & 39.2 & 36.7 \\
\hline \hline
\end{tabular}
\end{center}
\end{table}

\newpage 

\vspace*{5cm}

\newpage

\end{document}